\documentclass{cjaa}                    

\usepackage{graphicx}                   
\input{epsf.sty}                        
\input{psfig.sty}                       

\newcommand{\msun}{M$_\odot$}

\begin{document}
\bigskip

\def\hst{{\it HST}}
\def\ngst{{\it NGST}}
\def\jwst{{\it JWST}}
\def\acs{{\it ACS}}
\def\wfpc2{{\it WFPC2}}
\def\stis{{\it STIS}}
\def\nicmos{{\it NICMOS}}
\newcommand{\eg}{{\it e.g., }} 
\newcommand{\ie}{{\it i.e.}} 
\newcommand{\etal}{{\it et al.}} 
\newcommand{\lta}{{\>\rlap{\raise2pt\hbox{$<$}}\lower3pt\hbox{$\sim$}\>}}
\newcommand{\gta}{{\>\rlap{\raise2pt\hbox{$>$}}\lower3pt\hbox{$\sim$}\>}}

   \title{On the Reionization of the Universe 
}

 \author{Nino Panagia
      \inst{}\mailto{}
      }
   \offprints{Nino Panagia}                   
 \institute{Space Telescope Science Institute, 3700 San Martin
 Dr., Baltimore, MD 21218, USA\\
	   \email{panagia@stsci.edu}
          }


\abstract{  I present first principle model calculations for the 
reionization of the Universe, which provide simple diagnostics to 
identify the reionization sources. Within this frame, I also discuss the
results derived from the Hubble  Ultra-Deep Field observations, and the
possible need for future  observations with JWST to provide the final
answer.
   \keywords{cosmology: early universe 
   		--- cosmology: observations --- galaxies: evolution --- 
		galaxies: individual (HUDF 033238.7-274839.8) 
		}
   }

   \authorrunning{N.\ Panagia }            
   \titlerunning{Reionization of the Universe}  


   \maketitle
%
%
\section{Introduction}           
\label{sect:intro}

After the Big Bang  the gas in the Universe was
fully ionized and stayed such until the recombination rate exceeded
the ionization rate at $z\sim1,100$, \ie~when the temperature dropped below
$\sim3,000~K$.  Afterwards, the gas quickly became  completely neutral,
and the Universe entered its own {\it Dark Age} (Rees 1997).  We have to
wait until the birth of the first stars, galaxies or  condensed bodies,
formed  out of density fluctuations of the Universe, to have ionization
sources that were able to ionize matter in the Universe again.

Thus, the reionization of the intergalactic medium (IGM) is the last global
phase transition in the Universe, after which individuality prevailed
over collective behaviour. Actually,  since reionization
greatly reduced the opacity of the Universe to ionizing radiation, 
 it is likely  to  have influenced   the formation and
evolution of galaxies and other structures in the early Universe (see,
\eg~Loeb \& Barkana 2001). 

The presence of strong Lyman-$\alpha$ absorption in the spectra of
distant quasars indicates that reionization was completed by $z \approx
6$ (Becker \etal~2001, Fan \etal~2002), while the polarization of the
cosmic microwave background (CMB) radiation suggests it may have begun
much earlier, at $z \sim 17\pm 5$ (Spergel \etal~2003).

The ionizing radiation may be produced by {\it fusion}, \ie~ by early
type stars that evolve through  a series of nuclear burnings, or by {\it
gravitational energy}, \ie~ through accretion on black holes of various
sizes, such as QSOs, AGNs, or remnant BHs from massive star deaths (\eg
Barkana \& Loeb 2001).  
In the case of fusion, since about 7~MeV are released in H-burning, in
which 4 H atoms merge to form a He atom,  but only 13.6 eV are required
to ionize an H atom, it follows that a mass fraction of 8$\times
10^{-6}$ undergoing fusion is all is {\it strictly} needed to re-ionize
all hydrogen.  However, a realistic treatment of the process must take
into account the facts that only a fraction of the Lyman continuum
photons  may escape from any given source, and that the ionized gas may
recombine, and does so at a particularly high rate if  the IGM material
is clumped. These effects reduce the overall efficiency of the
ionization process. Another aspect to keep in mind is that reionization
is a statistical process, so that, especially in  the presence of strong
sources, the reionization of the Universe may have been a rather
inhomogeneous process, and observationally the ionization along
different lines of sight may be quite different.

Bearing these points in mind, in the following sections I shall quickly
review the efforts done in  recent years to clarify how reionization
occurs in the early Universe,  and which are the means to trace this
phenomenon in the cosmic evolution and to understand what is the nature
of the sources responsible for reionization.

\begin{figure} 
\centering
\includegraphics[height=9.cm]{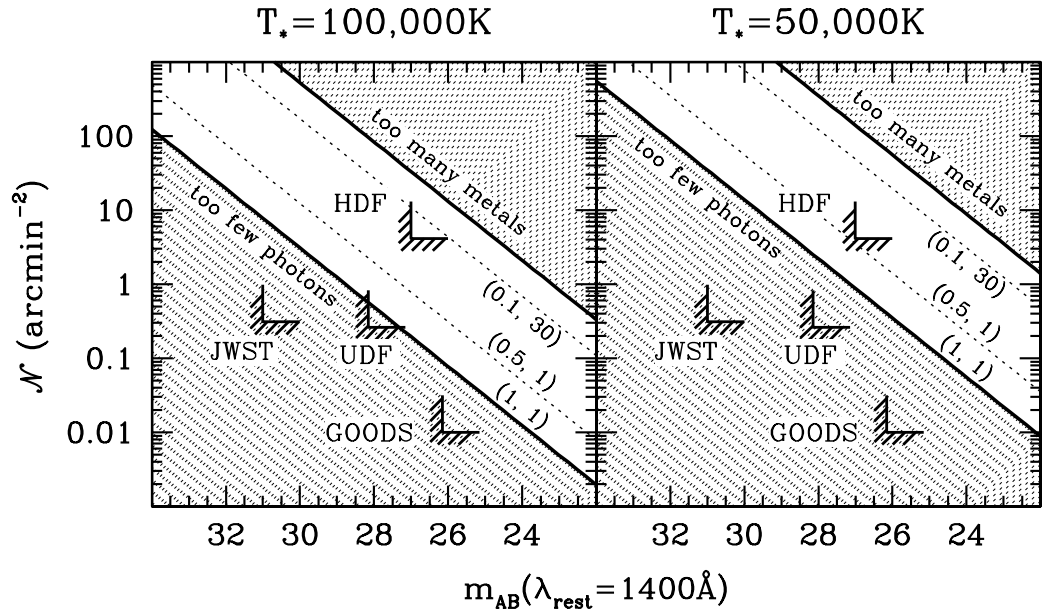}
\caption{The loci of surface density vs apparent AB magnitude for
identical reionization sources that are either Population III (left hand
panel) or Population II stars (right hand panel). In each panel, the
model curves are labelled with their $(f,C)$ values.  The L-shaped
markers delimit the areas probed by the GOODS survey, the HDFs, the HUDF,
and an ultra-deep survey with JWST [adapted from Stiavelli \etal~(2004a)]. 
\label{Fig.1}}  
\end{figure}

\section{First principle model calculations}

The theoretical efforts to understand and constrain the time evolution
of cosmic reionization sources  (\eg~Loeb \& Barkana 2001, and references
therein;  Stiavelli, Fall \& Panagia 2004a, hereafter referred to as
SFP) have to be confronted with observational identifications of the
ionization sources.  

\begin{figure} 
\centering 
\includegraphics[height=12cm]{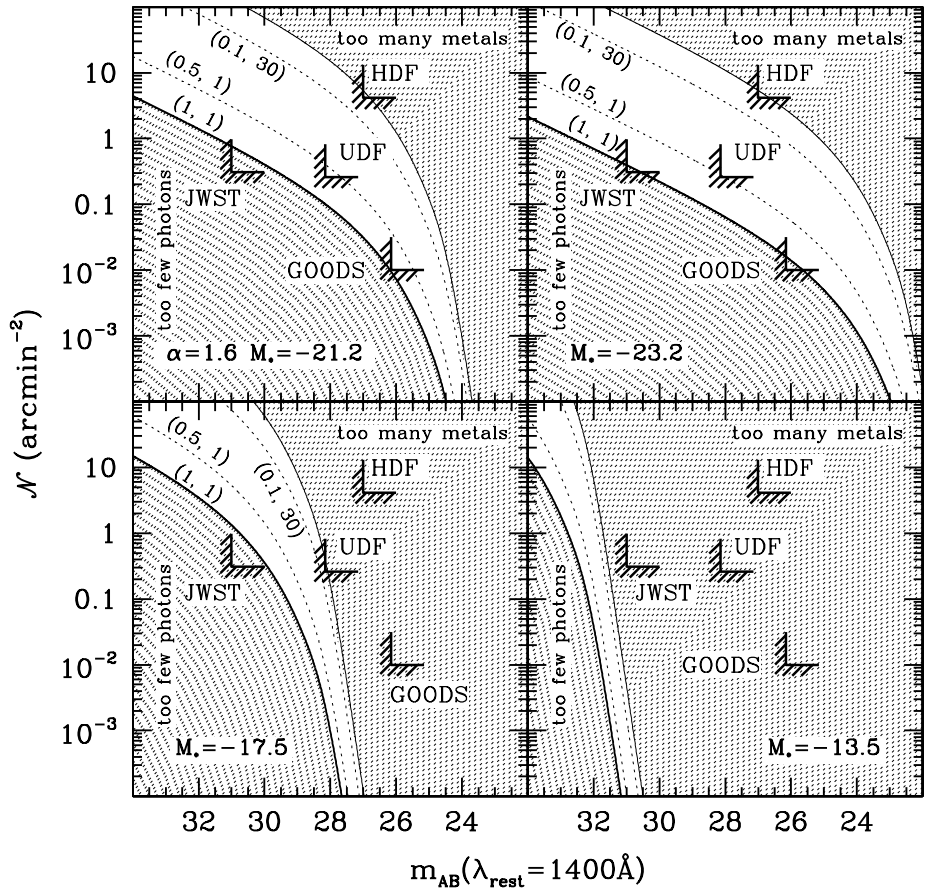}
\caption{The cumulative distribution of the surface density vs apparent
AB magnitude of reionization sources with luminosity functions with
different knees.   In each panel, the model curves are labelled with
their $(f,C)$ values.  The L-shaped markers delimit the areas probed by
the GOODS survey, the HDFs, the HUDF, and an ultra-deep survey with
JWST [adapted from Stiavelli \etal~(2004a)]. 
\label{fig2}} 
\end{figure}

In their seminal study SFP have considered the detectability of  the
sources responsible for the reionization of the Universe. The main idea is that
reionization places limits on the mean surface brightness of the
population of reionization sources. SFP defined a family of models
characterized by two parameters: the Lyman continuum escape fraction $f$
from the sources, and the clumpiness parameter $C$ of the intergalactic
medium, defined as  $C=<n{^2_H}>/<n_H>^2$.  The minimum surface
brightness model corresponds to a value of unity for both parameters. An
upper limit to the surface brightness is obtained by requiring that the
reionization sources do not overproduce heavy elements. SFP general
approach is applicable to most types of reionization sources, but in
specific numerical examples they focus on  Population III stars, because
these have very high effective temperatures and, therefore, are very
effective producers of ionizing UV photons ({\it e.g.,~}
Panagia \etal~2003). 
Figure 1 shows the loci of the mean surface brightness of {\it
identical} reionization sources  as a function of their observable AB
magnitude. The left panel refers to Population III sources with effective
temperature of $10^5$ K,  the right panel to Population II reionization
sources with effective temperature $5 \times 10^4$ K. In both panels,
the  lower solid line represents the minimum surface brightness model,
(1,1), while the upper solid line represents the global metallicity
constraint $Z\leq 0.01 Z_\odot$ at $z=6$. The  dotted lines
represent the (0.5, 1) and (0.1, 30) models.  The non-shaded area is
the only one accessible to reionization sources that do not overproduce
metals. The L-shaped markers delimit the quadrants ({i.e.,} the areas
above and to the right of the markers) probed by the GOODS/ACS survey
(Giavalisco \etal~2004), the HDF/HDFS NICMOS fields
(Thompson \etal~1999, Williams \etal~2000), 
the Hubble Ultra Deep Field (HUDF;
Beckwith \etal~2005)  and by a
hypothetical ultra-deep survey with JWST.

Figure 2 shows the expected cumulative surface density
distributions for reionization sources with a variety of   luminosity
functions. In each panel, the upper solid line represents the global
metallicity constraint. The lower solid line represents the minimum
surface brightness model, (1,1). The dotted lines give the
luminosity function for the (0.5,1) and the (0.1, 30) models. The
L-shaped markers delimit the quadrants probed by the GOODS/ACS survey,
the HDF and HDFS NICMOS fields, the HUDF, and an ultra-deep survey with
JWST, respectively.

From these results it appears that if reionization is caused by
UV-efficient, minimum surface brightness sources, the non-ionizing
continuum emission from reionization sources will be difficult to
detect before the advent of JWST.  On the other hand, if the sources of
reionization were not extremely hot Population III stars but cooler
Population II stars or AGNs, they would be brighter by 1-2 magnitudes
and thus they would be easier to detect.

\section{As Deep as Ever: the \acs~ Ultra-Deep Field}

The installation on HST of the new ACS camera in 2002 with its tenfold
increase of sensitivity opened  new possibilities for the exploration of
the Universe. A first project to exploit the new  capabilities for
cosmology was the GOODS Treasury program (Giavalisco \etal~2004) that
between August 2002 and May 2003  devoted almost 400 orbits of {\it HST}
time to observe with the ACS a total area of about 0.1  square degrees
in two fields, the Hubble Deep Field North (HDF-N) and the Chandra Deep
Field  South (CDF-S).  One of the primary scientific goals was the study
of the  evolution of star formation activity in galaxies and the very
assembly of galaxies up to redshifts  of about 6 (less than 1 billion
years past the Big Bang). 

While the GOODS project was focused on covering a relatively large area
to maximize the  statistical significance of their results, an even more
ambitious project has been implemented to  reach the ultimate frontiers
of the Universe with HST, the Hubble Ultra-Deep Field project (HUDF,
Beckwith \etal~2005). 

The HUDF consists of a single ultra-deep field (412 orbits in total: a
{\it million-second-long} exposure!))
within the Chandra Deep  Field South (CDFS) in the GOODS area. The
exposure time was divided among four filters, F435W (B),  F606W (V),
F775W (i'), and F850LP (z), to give approximately uniform limiting
AB-magnitudes near 31 for point sources. The pointing was selected so
as to avoid the gaps with  the lowest effective exposure times on the Chandra
ACIS image of the CDFS, and was designed to include  in the field both a
spectroscopically confirmed z=5.8 galaxy and a spectroscopically
confirmed  type Ia SN at z=1.3.  The {\it HST} observations begun on
September 24, 2003,  and continued through January 15, 2004. The reduced
ACS images and source catalogs were released on March 9, 2004, about
seven  weeks after the completion of all {\it HST} observations. The
resulting images are one magnitude deeper in the blue  and the visual,
and 1.5 magnitudes deeper in the red than the equivalent HDF exposures.
The  depth in the reddest filters is {optimized} for searching very red
objects -- like z=6 galaxies -- at the  detection limit of the
near-infrared image. 

The first results are presented in an archival paper (Beckwith \etal~
2005) that includes a catalog of more than 10,000 extended  objects, the
vast majority of which are galaxies.  Examination of the catalog for 
``dropout'' sources yields 312 B-dropouts, 79 V-dropouts, and 45
i-dropouts. A comparison of the limiting magnitudes and surface
brightnesses with crude estimates of galaxy properties in the early
Universe indicates that the field does contain a population of objects
at redshifts approaching 7.

In summary, the GOODS  project observations  (Giavalisco \etal~2004),
and the observations of the  Hubble Ultra-Deep Field (HUDF; Beckwith
\etal~2005) opened a window through which one may hope to recognize
distant sources that could provide a direct clue to the reionization
process.  The available  data have then extended the searches to $z\sim
6.5$ by  the drop-out techniques  (Dickinson \etal~2004, Giavalisco
\etal~2004, Bunker \etal~2004, Yan \& Windhorst 2004, Bouwens
\etal~2004b, Eyles \etal~2005) and narrow-band surveys  (Rhoads
\etal~2004, Stern \etal~ 2005, Taniguchi \etal~ 2005). Yet measures of
the luminosity functions remain uncertain (Yan \& Windhorst 2004, 
Bouwens \etal~2004a), and the identification of the reionization sources
is still just a possibility (\eg~Stiavelli, Fall \& Panagia 2004b). 
This is because, even with the exquisite images of the HUDF, it is hard
to identify and characterize sources at redshifts appreciably higher
than 6, which are crucially important to properly define and understand
the processes associated with reionization.  

\section{Detecting possible Reionization Sources}

The general method developed by SFP is applicable to a wide variety of
reionization sources. For specific predictions, however, it is
convenient to consider sources made either of metal-free Population~III
stars, with an effective temperature of $10^5$~K, or metal-weak
Population~II stars, with an effective temperature of $5\times 10^4$~K
(Stiavelli, Fall, \& Panagia 2004b). These effective temperatures are
nearly independent of the stellar initial mass function, provided only
that it extends beyond about 30~$M_{\odot}$. The metallicity dividing
these stellar populations is roughly $10^{-3}Z_{\odot}$.  Since
Population~III stars are hotter than Population~II stars, they produce
more ionizing photons for a given flux at longer wavelengths.  In this
sense, Population~III stars are also more efficient ionizers than active
galactic nuclei (AGN).

Figure~3 shows the expected cumulative surface density of reionization
sources as a function of their apparent AB magnitude in the non-ionizing
UV continuum at a rest-frame wavelength of 1400~\AA~  (Stiavelli, Fall,
\& Panagia 2004b). Here it is assumed that the comoving volume density
of the sources is constant over the range of redshifts $5.8 \lta z \lta
6.7$ spanned by $i$-band dropouts in the HUDF and GOODS (see below). The
luminosity function of the sources is assumed to have the Schechter
form, parameterized by its knee $M_*$ and slope $\alpha$. For reference,
the Lyman-break galaxies at $z = 3$ have $M_{*,1400} = -21.2$ and
$\alpha = 1.6$ (Steidel \etal~1999, Yan \etal 2002). These are the
parameters adopted for the top panels of Figure~3. The middle panels
have a brighter knee ($M_{*,1400} = -23.2$), and the bottom panels have
a steeper slope ($\alpha = 1.9$). The predictions in the left panels are
for Population~III stars; those in the right panels are for
Population~II stars. 

\begin{figure} 
\centering 
\includegraphics[height=14cm]{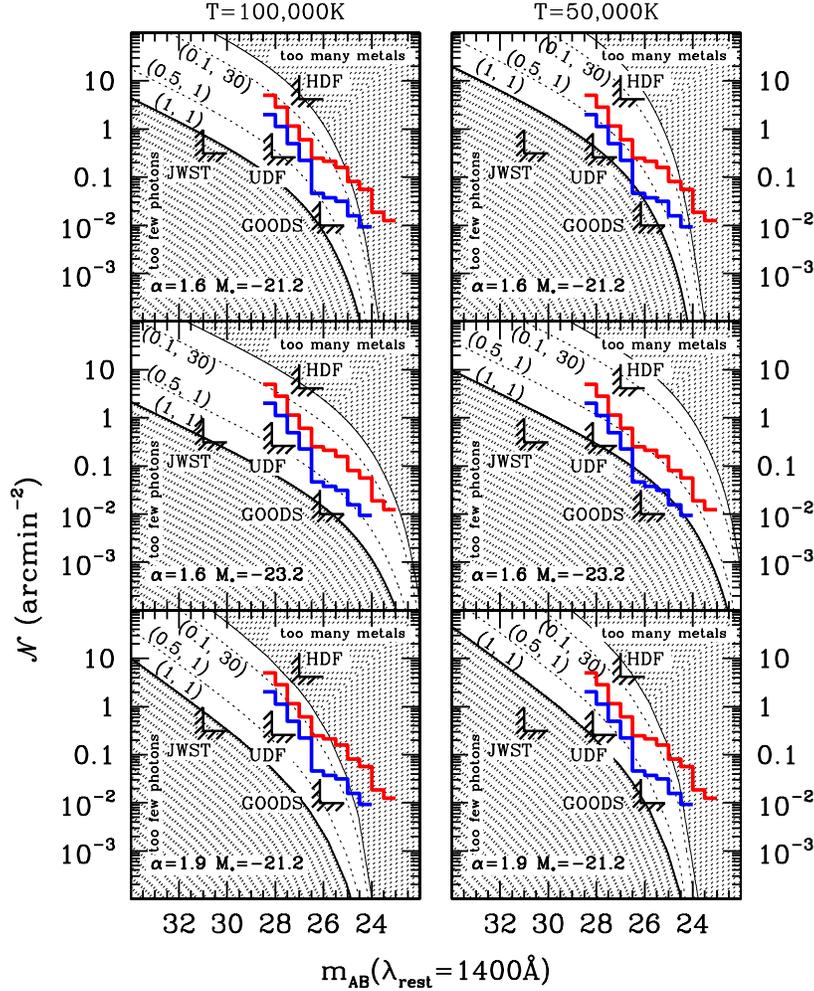}
\caption{Cumulative mean surface density vs apparent AB magnitude of
reionization sources with different luminosity functions and effective
temperatures. In each panel, the model curves are labelled with their
$(f,C)$ values. The thick stepped lines represent the observations from
the HUDF and GOODS with the two color cuts: $i-z\geq1.3$ (red lines) and
$i-z\geq2$ (blue lines).  The L-shaped markers delimit the areas probed
by the GOODS survey, the HDFs, the HUDF, and an ultra-deep survey with
JWST [adapted from Stiavelli \etal~(2004b)]. 
\label{fig2}} 
\end{figure}

The theoretical predictions (Stiavelli \etal~2004b) can be compared with
the observations from the HUDF and GOODS projects. The analysis is based
on the $i$-band dropouts ($z \approx 6$ galaxies) identified by Bunker
\etal~ (2004).  The magnitudes and colors of the objects were measured
in apertures 0.5 arcsec in diameter.  The HUDF catalog includes 53
objects with $i-z \geq 1.3$ and 22 with $i-z \geq 2.0$ down to a
limiting magnitude of $z=28.5$ at $S/N=8$, where the incompleteness is
only 2\%  (Table~1 of Bunker \etal~2004). Stiavelli \etal (2004b) 
analyze the $i-z \geq 1.3$ and $i-z \geq 2.0$ subsamples separately. 
The latter may miss some high-reshift objects, but it should be nearly
free of contamination by low-redshift objects.

Combining these HUDF observations with wider and shallower, but
otherwise similar, observations from the GOODS survey one obtains more
reliable estimates of the bright part of the surface density-magnitude
relation. In particular,   the GOODS observations were used brightward
of $z=26.5$ and the HUDF observations faintward of this magnitude.   The
resulting cumulative surface density-magnitude relation is shown by the
stepped lines in each panel of Figure~3.  The upper, red line refers to
the $i-z\geq1.3$ color cut, while the lower, blue line refers to the
more stringent $i-z\geq2.0$ color cut. One expects the true relation to
lie between the two stepped lines. Both of these relations appear
generally compatible with the predictions, especially when allowance is
made for the large statistical uncertainties at the bright ends. The
observed relation  extends into the shaded region at the bright end, but
we have checked that this does not violate the global metallicity
constraint, which pertains to the mean surface brightness when
integrated over all magnitudes. It appears that the integrated surface
brightness of the objects brighter than our limiting magnitude ($z_{\rm
lim} = 28.5$) is $\mu_{AB}=25.4$ and 26.7 mag arcsec$^{-2}$,
respectively, for the $i-z\geq1.3$ and $i-z\geq2$ subsamples. These are
consistent with the predicted minimum surface brightness $\mu_{AB}=28.8$
and 27.2 mag arcsec$^{-2}$ for reionization by Population~III and
Population~II sources, respectively, and a redshift interval $\Delta
z=1$ (SFP).

\section{HUDF-JD2, a powerful source at z$> 6.5$}

\begin{figure}[t!] 
\centerline{\epsfxsize=11cm\epsfbox{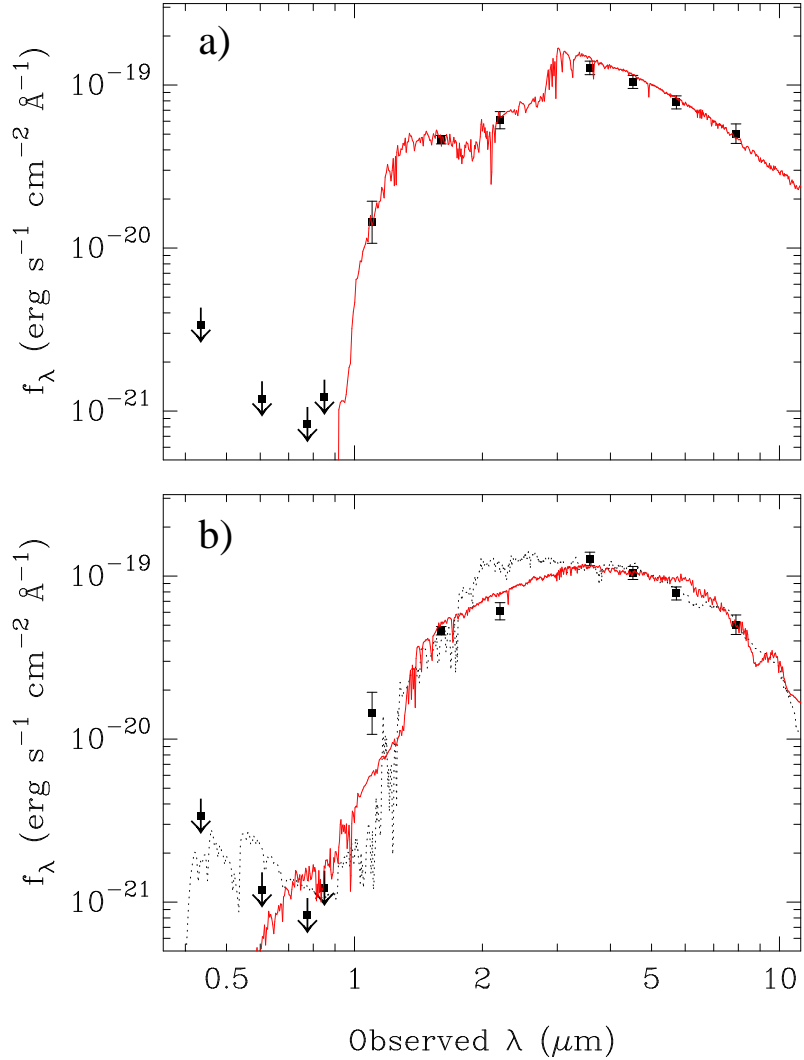}}
\caption{The observed SED of HUDF-JD2 is compared with theoretical
models at different redshifts (Mobasher \etal 2005). It is apparent that
the high redshift solution (upper panel: z$\sim 6.5$) provides an
excellent fit to the observations.Lower redshift best-fit solutions
(lower panel): the red curve is a ``dusty starburst" model  at z=2.5, the
dotted curve  is an old population model at z=3.4. They both provide
much poorer fits. } 
\end{figure}

In order to provide the multi-waveband information needed to search for
galaxies at even higher redshifts   and to explore their nature
individually, one can use combined ultra-deep images of faint  galaxies
taken by both the  {\it Hubble Space Telescope} and the {\it Spitzer
Space Telescope}.  In particular,  the HUDF  observations (Beckwith
\etal~2005, Thompson \etal~2005), which represent  the deepest images
of the Universe at optical  ($B$, $V$, $i$, and $z$ bands; HST/ACS) and
near-infrared ($J$, $H$; HST/NICMOS) wavelengths, combined with the VLT
and Spitzer/IRAC observations obtained by the GOODS project at NIR
($K_s$; ESO/VLT) and mid-infrared (3.6-24 $\mu$m;  Spitzer/IRAC+MIPS)
wavelengths (B. Vandame \etal~ 2005, in preparation, I. Labb\'e
\etal~2005, in preparation, M. Dickinson \etal~2005, in preparation),
are well suited to accomplish this aim.

Analyzing those data, Mobasher \etal~(2005)  searched for  very red
galaxies with $(J-H)_{AB}>1.3$ and no detection at wavelengths shorter
than $J$ (\ie~$J-$band dropouts). They found one source that  satisfies
the $J-$band dropout criteria, is well detected in the K-band (ISAAC)
and in all four IRAC channels (3.6, 4.5, 5.8 and 8 $\mu$m) and appears
to be at a redshift $z\sim6.5$ or higher: HUDF033238.7-274839.8, 
hereafter referred to as HUDF-JD2.

The presence of a clear Balmer break in the observed SED of HUDF-JD2
(see Fig.~4) confirms its high redshift identification and  reveals a
post-starburst population.   From their best-fit models,  Mobasher
\etal~(2005) derive a photometric redshift of $z\sim6.5$ and a
bolometric luminosity of L$_{\mathrm{bol}} = 1 \times 10^{12}$
L$_{\odot}$ (for a cosmology with $H_0=70$ km s$^{-1}$ Mpc$^{-1}$,
$\Omega_{m}=0.3$ and $\Omega_{\Lambda}=0.7$).   They also estimate the
mass  in stars is M$_* \simeq 6 \times 10^{11}$ M$_{\odot}$ and that
the stars were formed at redshifts higher than z$\sim$9. Moreover,
HUDF-JD2 had to form the bulk of its stars very rapidly, on time scales
$\leq 100$ Myr, so that  the subsequent evolution was essentially passive. 
A stringent upper limit to the starburst  age is set by the photometric
redshift $z\simeq 6.5$ of HUDF-JD2, which corresponds to a time when the
Universe was only 830 Myr old.  Adopting this prior, Panagia
\etal~(2005) infer that the age of the stellar population in HUDF-JD2 is
likely to be bracketed between 350 and 650 Myr.  These ages correspond
to redshifts of galaxy formation between 10 and 20.
The estimated reddening is quite modest, with $E(B-V)\leq 0.06$ at 95\%\
confidence limit. The best-fit model SED also provides estimates of the
overall metallicity of the galaxy, which is  bracketed within the
interval $Z=0.02-0.004$, \ie~between $Z_\odot$ and $Z_\odot/5$, so that 
a fiducial value of  $Z=0.008=0.4Z_\odot$ has been adopted.

\section{IMPLICATIONS FOR REIONIZATION}

Reionization is a process that depends on the UV output of  galaxies
integrated over the  time interval when they are active UV sources.   In
this respect, HUDF-JD2 is a promising reionization source in that, 
according to the models, its integrated Lyman continuum photon flux is
about $4\times10^{72}$ photons  that had to ionize the neutral hydrogen
gas within a comoving volume of about $14-18\times10^4$ Mpc$^3$, the
exact volume depending on the specific redshift (within the range
$z\sim9-15$; see below) at which H was fully ionized.   For a comoving hydrogen
density of $6.2\times10^{66}$ atoms~Mpc$^{-3}$ (\eg~Spergel \etal~2003),
it appears that the produced Lyman continuum photons outnumber the  H
atoms by almost an order of magnitude and that, therefore, a complete
reionization is a real possibility (Panagia \etal~2005).  
However, as pointed out in the introduction, a realistic treatment of
this  process must take into account the facts that only a fraction
of the Lyman continuum photons may escape from the source, and that
the ionized gas may recombine, especially if  the IGM material is
clumped, because both effects reduce the overall efficiency of
the ionization process.

\begin{figure}[t!] 
\centerline{\epsfxsize=10cm\epsfbox{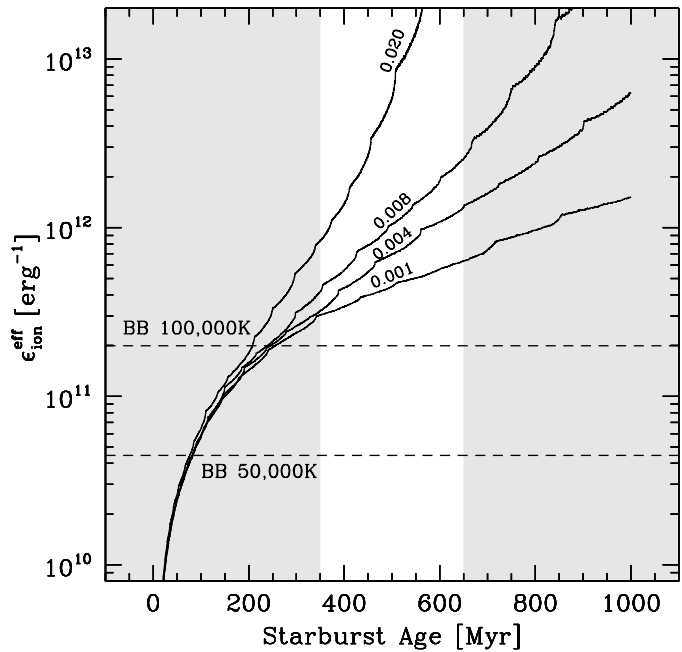}}
\caption{Effective ionization parameter $\epsilon{^{eff}_{ion}} =
<N_{Lyc}>/(\nu L_\nu)_{1400}$ as a function of starburst age  for
different metallicities (indicated as metal abundances by mass).  For
comparison, the values of the ionization parameters for black-bodies
with temperatures of 50,000 and 100,000~K are also shown. The unshaded
area corresponds to the interval within which the age of HUDF-JD2 is
likely to lie [adapted from Panagia \etal~(2005)].}
\end{figure}

Panagia \etal~ (2005) have approached this problem by using the
extensive set of model calculations made by SFP.  However, all cases
studied by SFP are based on the implicit assumption of a ``snapshot" of
a statistical assembly of cosmic sources, which may have  formed at
different epochs and were  efficient UV radiation emitters over
different time intervals, but, {\it on average}, constituted a steady
supply of ionizing photons over the chosen redshift interval.  The
situation of HUDF-JD2 is different. HUDF-JD2 is  a single, powerful
source for which both the birth time and  subsequent evolution are
constrained by the observed SED. In particular,  we know that the galaxy
formed stars and was very bright in its first 100 Myr of age, and
essentially was evolving passively afterwards, thus declining in
luminosity with time.  The effects of this type of evolution can be
included by adopting an effective ionization parameter
$\epsilon{^{eff}_{ion}} = <N_{Lyc}>/(\nu L_\nu)_{1400}$, in which the
ionizing photon flux is averaged over the 100 Myr of starburst activity
whereas the instantaneous source luminosity is declining as the galaxy
is aging (see Figure~5).  It turns out that the effective ionization
parameter $\epsilon{^{eff}_{ion}}$ appropriate  for HUDF-JD2 is a factor
of about 5 higher than it is for a 100,000~K black-body. This is because
the integrated HUDF-JD2 output in ionizing radiation is high whereas its
presently {\it observed} UV flux has considerably decreased as a
consequence of HUDF-JD2 passive evolution. 

\begin{figure}[t!] 
\centerline{\epsfxsize=10cm\epsfbox{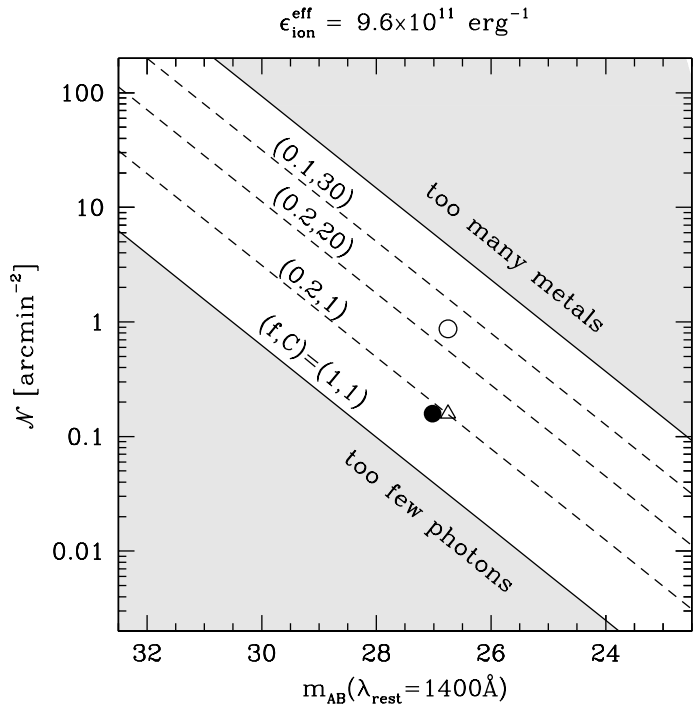}}
\caption{Surface density of reionization sources as a function of the
apparent AB magnitude for sources with ionization parameter
$\epsilon{^{eff}_{ion}} = 9.6\times 10^{11}$~erg$^{-1}$. The lower solid
line represents the minimum surface brightness model, $(f,C)=(1,1)$, 
while the upper solid line represents the global metallicity constraint
$Z < 0.01 Z_\odot$ at z = 6.5 (see SFP). The dashed lines represent
models with  lower escape fraction of ionizing photons and higher
clumping factor, \ie $(f,C)=(0.2,1)$, $(0.2,20)$, and $(0.1,30)$,
respectively. The observed position of  HUDF-JD2 is denoted with a dot, 
whereas the triangle and the circle denote the positions after
correction for dust extinction and after allowance for unseen
companions, respectively  [adapted from Panagia
\etal~(2005)].}
\end{figure}
 
The corresponding constraints to the surface density  of possible
reionization sources as a function of the apparent AB magnitude are
shown in Figure 6.  We see that HUDF-JD2 {\it as observed}, having a
J-band magnitude of $\sim27$ and being one source in a field of 6.25
square-arcminutes (\ie~ the HUDF area covered with NICMOS observations),
lies well above the locus of minimum surface brightness. It appears
that  HUDF-JD2 by itself could be responsible for the reionization of
the IGM in its region of Universe if the escape fraction of ionizing
radiation is higher than $\sim0.25$ and the IGM gas is uniformly
distributed, or conversely if the clumping factor is as high as 20 but
the bulk of the ionizing radiation is able to escape from the galaxy. 
Allowance for some extinction correction and, more importantly, for the
presence of unseen fainter companions, would shift the point up by a
factor of 5-6, thus making it capable of ionizing its region entirely
even for a low escape fraction, $f\sim0.2$,  {\it and} high clumping
factor, $C\sim20$ (see Panagia \etal~2005 for details).

It is interesting to note that the four $i'$-dropout sources detected
and studied by Eyles \etal~ (2005) in the HST/ACS GOODS images of the
Chandra Deep Field South, may indeed be representatives of  the ``second
brightest" galaxies in the ladder of the luminosity function. Actually,
their measured magnitudes are about 1.5-2 mag fainter than measured for
HUD-JD2.  All of these sources are found to have z$\sim5.8$, stellar
masses  $2-4\times 10^{10}$\msun, ages of $250-650$ Myr and implied
formation redshifts $z_f\sim7.5-13.5$ (Eyles \etal~ 2005).  The presence
of HUD-JD2 together with these additional bright sources suggest that the
reionization of the Universe may have been dominated by massive galaxies that,
rather unexpectedly, were able to form  during the early evolution of
the Universe.

\section{DISCUSSION AND CONCLUSIONS}

It appears that HUDF-JD2  may indeed  be capable of reionizing its
portion of IGM (\ie~the gas within a volume  defined by the subtended
solid angle of the HUDF-NICMOS field and the redshift of complete
reionization, z$_2\sim9-15$),  possibly with the help of unseen fainter
companions, starting the process at a  redshift as high as z$_1\simeq 15
\pm 5$. Regardless of whether HUDF-JD2 did ionize its cell of Universe 
{\it entirely}, it is certain that it had the power to appreciably
ionize it at high redshifts($z\sim10-20$), so as to account for WMAP
findings ($\tau_e=0.17\pm0.06$ at z$=17\pm5$; Spergel \etal~ 2003).

Although the presence of a cluster of galaxies and the assumptions about
their properties are somewhat speculative, we would like to stress that
these are {\it predictions} that will be possible to confirm or reject
once the James Webb Space Telescope (JWST; \eg~Stiavelli \etal~ 2004c),
currently due to be launched in mid-2013, will come into  operation.  
According to the model calculations by Panagia \etal~(2003, and in
preparation), sources like HUDF-JD2 and its analogs at higher redshifts
will be easy to study spectroscopically because in their  active star
formation phases, they will have intrinsically high luminosities
providing fluxes as high as 150~nJy in the near and mid infrared. 
Fluxes at these levels are expected to be detectable with the
JWST-NIRSpec spectrograph at a resolution of $\sim1000$ with a S/N ratio
of 10 in 10$^5$ seconds exposures (\eg Panagia \etal~2003, Stiavelli et
al. 2004c).   Therefore, with the advent of JWST it will be possible to
fully characterize these early Universe sources, so as to clarify
conclusively the processes of galaxy formation and the reionization of
the Universe.

\end{document}